\documentstyle[aps,multicol,epsf,rotate]{revtex}

\newcommand{\be}{\begin{equation}}
\newcommand{\ee}{\end{equation}}
\newcommand{\bdm}{\begin{displaymath}}
\newcommand{\edm}{\end{displaymath}}
\newcommand{\ba}{\begin{eqnarray}}
\newcommand{\ea}{\end{eqnarray}}

\begin{document}

\draft
%\twocolumn[\hsize\textwidth\columnwidth\hsize\csname @twocolumnfalse\endcsname

\title{
Energy resolved supercurrent between two superconductors 
} 
\author{S.-K. Yip}
\address{ 
Department of Physics \& Astronomy, 
           Northwestern University, Evanston, Illinois 60208, U. S. A. \\
and\\
Department of Physics, {\AA}bo Akademi, 
           Porthansgatan 3, 20500 {\AA}bo, Finland} 
\date{\today}

\maketitle

\begin{abstract}
{\small

In this paper I study the energy resolved supercurrent
of a junction consisting of a dirty normal metal between
two superconductors.  I also consider a cross geometry
with two additional arms connecting the above mentioned
junction with two normal reservoirs at equal and
opposite voltages.  The dependence of the 
supercurrent between the two superconductors
on the applied voltages is studied.

 }
\end{abstract}
\date{\today}
\vspace*{0.2 cm}

\begin{multicols}{2}

The proximity effect between a normal metal and a superconductor
has been discussed long time ago \cite{deGennes}.  Often
it is simply described by a spatial dependent 
pairing correlation function $\Psi$ which decays from the
superconductor to the normal metal. 
However, this description is too crude to provide
a proper understanding of the phenomena observed
at low temperatures in the mesoscopic systems 
which can nowadays be prepared in the laboratories.
For example the detailed description of the
 energy dependence of the effective
barrier conductance and 
diffusion coefficient 
\cite{zaitsev90,volkov92a,volkov92b,volkov93,volkov94,zaitsev94,yip95}
is crucial in understanding
the behavior of the observed conductance between
  a normal metal (N) and a superconductor (S)
at low voltages and temperatures.
\cite{courtois96,charlat96,hartog96,poirer97}

In this paper we study the spectral current density
\cite{rainer96} (see also \cite{bardeen68}) of a
quasi-one dimensional SNS junction in the dirty
limit.  This quantity (or, more precisely,
the angular average of the one defined in \cite{rainer96}) is
defined as, at energy $\epsilon$ and position $x$,

\be
N_J (\epsilon, x) = < \hat p_x \ N (\hat p, \epsilon, x) >
\label{defNJ}
\ee

\noindent where $  N (\hat p, \epsilon, x) $ is the density of
states for momentum direction
$\hat p$ at energy $\epsilon$ and position $x$.
The angular brackets denote angular average.
This quantity is thus the density of states
weighted by a factor proportional
to the  current that each state carries
(in a certain direction, here $\hat x$),
and thus may also be appropriately
referred to as the current-carrying density of states.
This is obviously a useful quantity.
For example at equilibrium, the (number) supercurrent $J_s$
can be written as

\be
J_s = - 2 v_f \int {d \epsilon \over 2} N_J ( \epsilon, x) 
                h_0 (\epsilon)
\label{Js}
\ee

\noindent where $ h_0 (\epsilon) = {\rm tanh} {\epsilon \over 2 T} $
and $v_f$ is the fermi velocity.  The factor
of $2$ includes the contribution from the two spin directions.
One convenient
way to interprete this formula \cite{bardeen68,rainer96}
(see also \cite{yip92,xu95})
is to rewrite $ h_0 = ( 1 - 2 n)$ where $n(\epsilon)$,
the occupation number, 
is given by the Fermi function at equilibrium. 
For example at $T=0$ eqn (\ref{Js}) can be re-written as
(using the symmetry $N_J (\epsilon) = - N_J ( - \epsilon)$ )

\be
J_s =  2 v_f \int_{-\infty}^0 {d \epsilon} N_J ( \epsilon) 
\label{Js0}
\ee

\noindent and thus can be interpreted as the current
due to the occupation of negative energy states.
This can also be regarded as
 the diamagnetic response of the superconductor if
one considers the $T=0$ state as one containing no
quasiparticles.
Similarly at finite temperature

\ba
J_s (T) &=& J_s(T=0) +          \nonumber              \\
     & &      2 v_f \int_{-\infty}^{\infty} {d \epsilon} N_J ( \epsilon) 
                 ( n (\epsilon, T) - n (\epsilon, T=0) )
\label{JsT}
\ea

\noindent and can be interpreted as the sum of the diamagnetic current
and the correction due to the thermal redistribution
of quasiparticles.
In particular an important source of the decrease of
the supercurrent as the temperature increases
is due to the thermal excitations of quasiparticles
from $\epsilon < 0$ to $ \epsilon > 0$  states, which
carry opposite current.

In the dirty limit, on which this paper will concentrate,
$N_J$ can be obtained from (see Appendix for details)

\be
N_J (\epsilon, x) = - {N_f l \over 6} Q (\epsilon, x)
\label{NJQ}
\ee

\noindent where $N_f$ is the density of states in the normal state,
 $l$ is the mean free path, and $Q$ is
given by

\be
Q \equiv {1 \over 4 \pi^2} {\rm Tr} 
            [ \tau_3 ( \hat g^R \partial  \hat g^R -
                         \hat g^A \partial  \hat g^A) ]
\label{defQ}
\ee

\noindent Here $  \hat g^{R,A}$ are the angular averaged of the
retarded and advanced components of the quasiclassical
Green's function. $\partial$ represents spatial derivative.
The equilibrium (number) supercurrent is thus given by

\be
J_s = { N_f D \over 2} \int d\epsilon Q h_0(\epsilon)
\label{JsQ}
\ee

\noindent where $D \equiv v_f l /3$ is the diffusion
coefficient.  For an
SNS junction with no electron-electron or
electron-phonon interaction in the N region,
$Q$ is independent of the position $x$ along the 
junction within that region.

The behavior of $Q$ is easiest to understand in the limit
of very short junction ( $E_D \equiv D/L^2 << \Delta$,
here $L$ is the length of the junction and $\Delta$
is the superconducting gap)
and small phase difference  $\chi$.
In this case $Q$ should be the same as that of a bulk
superconductor under a small phase gradient.  The response
of a dirty superconductor to a phase gradient or
an external vector potential is well-known \cite{abrikosov}.
In this case one can show that the entire contribution
to the  supercurrent arises from states at $ \epsilon = \Delta$,
{\it i.e.} $ Q \propto \delta (\epsilon - \Delta)$.  In contrast
the ordinary density of states is given by $N(\epsilon) =
N_f { \vert \epsilon \vert \over \sqrt{ \epsilon^2 - \Delta^2} }$.
Under a small phase gradient, the gap for quasiparticle excitations
persists and in particular there is no contribution to
$Q$ for energies within this gap.

An energy gap $\epsilon_g$ ($< \Delta$)
exists in general also in an SNS junction
(except phase difference $\chi = \pi$).  This gap
has been studied before in related situations
\cite{golubov88,belzig96,golubov97}.  Associated with the existence of
this (phase dependent) gap is a relatively rapid change of $\hat g$
as a function of energy (and  phase difference).
This has made the numerical calculation somewhat difficult.
For convenience I will thus mostly concentrate
on results where a small pair-breaking term $\gamma$ has been
included in the self-energy (see Appendix). 
$\gamma$ is usually chosen to
be $0.05 \Delta$, though occasionally results for
$\gamma = 0$ will  also be shown for comparison.

\begin{figure}
\centerline{ 
	\epsfysize=0.4\textwidth \rotate[r]{
	\epsfbox{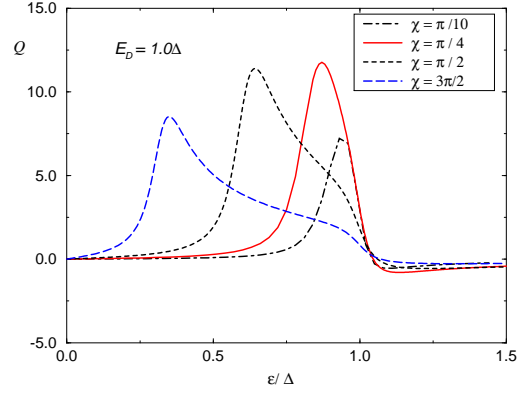} }
}
\vskip 0.5 cm
\begin{minipage}{0.45\textwidth}
\caption{
$Q$ (in units of $1/L$)
 for a short junction. $E_D = 1.0 \Delta$. $\gamma = 0.05 \Delta$ }
\label{fig:q-short}
\end{minipage}

\end{figure}

The behavior of $Q$ for a relatively short junction is as
shown in Fig. \ref{fig:q-short}.  At small phase
differences $Q$ is large only for $\epsilon$ near
$\Delta$.  If $\gamma$ were zero then
$Q$ would vanish for $\epsilon$ below a minigap $\epsilon_g$.
As the phase difference increases, the minigap decreases.
Correspondingly the region of energy where $Q$ is finite also
moves down in energy, though it remains large
in an energy region up to $\approx \Delta$.

%%%%%%%%%%%%%%%%%%%%%%%%%%%%%%%%%%%%%%%%%%%%%%%%%%%%%%%%%%%%%%%%%%%%%%

%%%%%%%%%%%%%%%%%%%%%%%%%%%%%%%%%%%%%%%%%%%%%%%%%%%%%%%%%%%%%%%%

\begin{figure}
\centerline{ 
	\epsfysize=0.4\textwidth \rotate[r]{
	\epsfbox{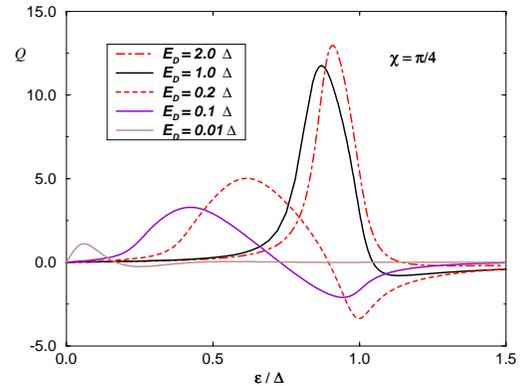} }
}
\vskip 0.5 cm
\begin{minipage}{0.45\textwidth}
\caption[]{
$Q$ for $\chi= \pi/4$ as a function of decreasing
$E_D$. $\gamma = 0.05 \Delta$ }
\label{fig:qe}
\end{minipage}
%\label{fig:qe}
\end{figure}

For longer junctions, {\it i.e.} $ L >> \sqrt{D/\Delta}$ or
equivalently $E_D << \Delta$, 
 the behavior is somewhat different.
At a given phase difference, the main region of energy where
$Q$ is significant is no longer of order $\Delta$.  
An example for this evolution as a function of increasing
length is as shown in Fig \ref{fig:qe}.  For a given
phase difference, the energy where $Q$ peaks shifts 
down in energy relative to $\Delta$ as $L$ lengthens.
  This itself may not be surprising, and can be understood by
analogy with the behavior of energy levels under a
change in boundary condition in the normal state.

%%%%%%%%%%%%%%%%%%%%%%%%%%%%%%%%%%%%%%%%%%%%%%%%%%%%%%%%%%%%%%%

%%%%%%%%%%%%%%%%%%%%%%%%%%%%%%%%%%%%%%%%%%%%%%%%%%%%%%%%%%%%%%

\begin{figure}
\centerline{ 
	\epsfysize=0.4\textwidth \rotate[r]{
	\epsfbox{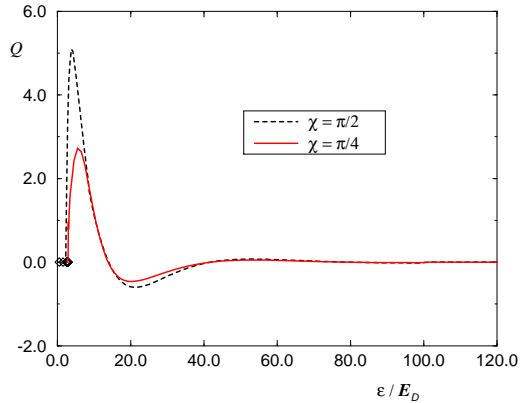} }
}
\vskip 0.5 cm
\begin{minipage}{0.45\textwidth}
\caption[]{
$Q$ for a long junction. $\Delta = 100 E_D$.
 This result is for $\gamma = 0$}
\label{fig:q-long}
\end{minipage}
%\label{fig:q-long}
\end{figure}

The more interesting feature is that a negative
dip in $Q$  appears at higher energies as the
junction lengthens.  For very long junctions,
both the peak and the dip of $Q$  move to
energies of order (a few tens times)  $E_D$, with
almost no features left near $\Delta$
(Fig. \ref{fig:q-long}).  This negative dip has been
speculated to exist recently \cite{argaman97}

%%%%%%%%%%%%%%%%%%%%%%%%%%%%%%%%%%%%%%%%%%%%%%%%%%%%%%%%%%%%%%%

%%%%%%%%%%%%%%%%%%%%%%%%%%%%%%%%%%%%%%%%%%%%%%%%%%%%%%%%%%%%%%%%

\begin{figure}
\centerline{ 
	\epsfysize=0.4\textwidth \rotate[r]{
	\epsfbox{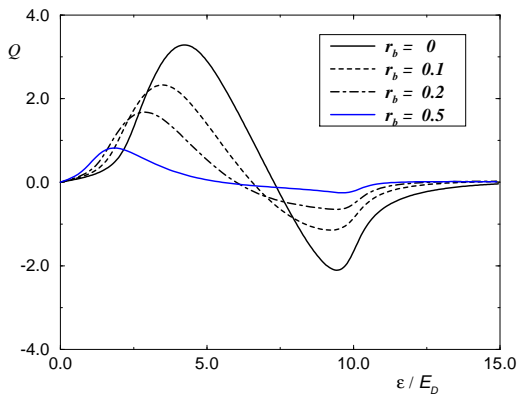} }
}
\vskip 0.5 cm
\begin{minipage}{0.45\textwidth}
\caption[]{
$Q$ for $\chi= \pi/4$, $\Delta = 10 E_D$ 
as a function of increasing
$r_b$, the ratio of the barrier
resistance $R_b$ to that of the normal metal, 
{\it i.e.} $r_b \equiv R_b  ( 2 N_f D S /L)$. 
Here $S$ is the area.
$\gamma = 0.05 \Delta$. }
\label{fig:qrb}
\end{minipage}
%\label{fig:qrb}
\end{figure}

In the above I have assumed that the contacts
between the normal metal N and the superconducting
reservoirs S are perfect.  If potential barriers
exist between the N and S regions, then $Q$ decreases in magnitude,
with a corresponding decrease in the energy 
where $Q$ peaks.  The features discussed above
survives for moderate barrier resistance $R_b$ between
N and S.  An example of how $Q$ evolves as $R_b$ increases
is as shown in Fig. \ref{fig:qrb}.

%%%%%%%%%%%%%%%%%%%%%%%%%%%%%%%%%%%%%%%%%%%%%%%%%%%%%%%%%%%%%%%

%%%%%%%%%%%%%%%%%%%%%%%%%%%%%%%%%%%%%%%%%%%%%%%%%%%%%%%%%%%%%%

\begin{figure}
\centerline{ 
	\epsfysize=0.4\textwidth 
\rotate[r]
{\epsfbox{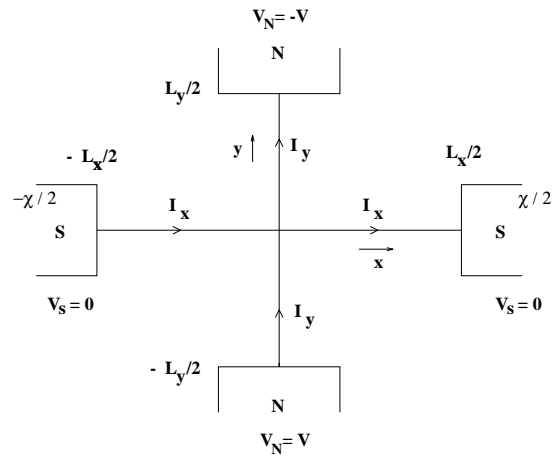} }
}
\begin{minipage}{0.45\textwidth}
\caption{
The cross geometry. All 'wires' connecting the reservoirs  are assumed to
be quasi-one-dimensional.}
\label{fig:cross}
\end{minipage}
\end{figure}

From the ideas presented above obviously one can affect
the current flowing between the two superconducting reservoirs
by changing the occupation of the quasiparticle states.
Temperature is an obvious candidate. This gives
the well-known reduction of the supercurrent
as a function of increasing temperature. 
%(see \cite{wilhelm96} for a recent study of this 
%temperature dependence).
An alternative way is to create a non-equilibrium situation
\cite{argaman97}.  Here I shall consider a steady state
situation with the advantage that it is easy to analyze.
The set-up is shown schematically in Fig \ref{fig:cross}.
 Geometries  closely related to this has been
studied before 
\cite{volkov95,nazarov96,stoof96,volkov96a,volkov96b,volkov97}. 
 However, these
references have concentrated on different arrangement
of voltages and/or other measureable quantities. 
Here I consider the case where the superconductors
are at equal voltages, chosen to be zero.
The normal reservoirs are at equal and opposite
voltages $V_N = \pm V$.  I shall study the
dependence of the current between the
superconducting reservoirs as a function of $V$.

%%%%%%%%%%%%%%%%%%%%%%%%%%%%%%%%%%%%%%%%%%%%%%%%%%%%%%%%%%%%%%%%%

%%%%%%%%%%%%%%%%%%%%%%%%%%%%%%%%%%%%%%%%%%%%%%%%%%%%%%%%%%%%%%%

\begin{figure}
\centerline{ 
	\epsfysize=0.4\textwidth 
\rotate[r]
{\epsfbox{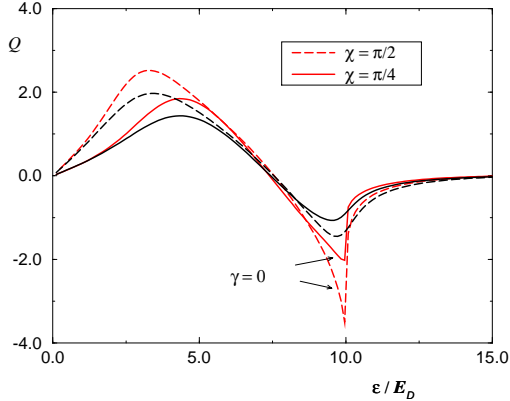} }
}
\vskip 0.5 cm
\begin{minipage}{0.45\textwidth}
\caption[]{ The quantity $Q$ (in units of $L^{-1}$) for the 
cross geometry  of Fig. 5.
$\Delta = 10 E_D$.
$\gamma = 0.05 \Delta$.  Results for $\gamma = 0$ are
also shown for comparison. }
\label{fig:crossq}
\end{minipage}

\end{figure}

First we should note that the presence of the
side arms connected to the normal
reservoirs affect the behavior of $Q$ via
the proximity effect.
In order to facilitate later discussion,
I plotted the quantity $Q$ for this spatial
geometry for the case $\Delta = 10 E_D$
for two phase differences
in Fig. \ref{fig:crossq}. 
In this example I have assumed that the arms
between the normal metal and the 
superconductor are symmetric and
of equal length ($L_x = L_y = L$ in Fig \ref{fig:cross} )
and area $S$. $Q$ is
finite only for the $x$ arms connecting the
superconducting reservoirs, and is constant along them.
Compared with the case without the side arms (Fig. \ref{fig:qe}),
we see that the behavior of $Q$ is somewhat different in
the energy region $\epsilon < E_D$.
This is because there is now no energy gap for quasiparticle
excitations for any position within the N region on the cross,
even for $\gamma = 0$.
However, for $\epsilon > E_D$ the qualitative behavior of $Q$
is almost the same as in the case without
the side arms, except an overall reduction in magnitude.
\cite{magnitude}
In particular the sign change of $Q$ remains.

%%%%%%%%%%%%%%%%%%%%%%%%%%%%%%%%%%%%%%%%%%%%%%%%%%%%%%%%%%%%%%%%%

%of Fig. \ref{fig:cross}. 
%%%%%%%%%%%%%%%%%%%%%%%%%%%%%%%%%%%%%%%%%%%%%%%%%%%%%%%%%%%%%%%

Obviously if $V = 0$ a supercurrent $I_s$ only flows
between the superconducting reservoirs, whereas 
there is no current flowing in or out
of the normal reservoirs.
 At $V \ne 0$ current is in general finite
at any position on the two arms. 
I shall denote the currents as $I_x$ and $I_y$.
 Neither
$I_x$ nor $I_y$ are position dependent;
moreover, the current flowing in and out
of the normal reservoirs are equal on
the one hand and those of 
the superconducting reservoirs 
equal on the other. 
One can therefore regard the current 
$I_x$ ($I_y$) as simply flowing between
the superconducting (normal) reservoirs.
I shall thus continue to call $I_x$ the
supercurrent $I_s$.   I  shall
consider how this $I_s$ is modulated by
the voltage $V$.
All results presented below are for $T=0$.

\begin{figure}
\centerline{ 
	\epsfysize=0.4\textwidth 
\rotate[r]
{\epsfbox{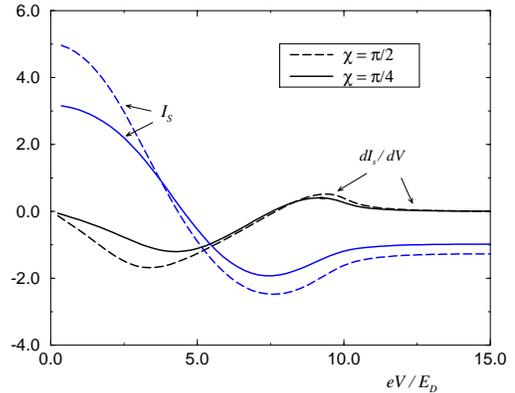} }
}
\vskip 0.5 cm
\begin{minipage}{0.45\textwidth}
\caption[]{ $dI_s/dV$ versus $V$ for the
cross geometry of Fig. 5. 
$E_D = 0.1 \Delta$, $\gamma = 0.05 \Delta$.
Also shown are $I_s$ as functions of $V$. 
Energies ($eV$) are in units of $E_D$ and $I_s$
 is in units of $N_f E_D D S L^{-1}$.}
\label{fig:IV}
\end{minipage}
\end{figure}

I shall concentrate on an example in the  most interesting
regime,  where $\Delta \sim 10 E_D$.  The result
for $d I_s / dV$ at $E_D = 0.1 \Delta$ is as
shown in Fig. \ref{fig:IV}.  In this parameter
range $d I_s / dV$ at a voltage $V$ is approximately equal to
$- (N_f D S ) Q $ at the corresponding energy $\epsilon = e V$.
({\it c.f.} the corresponding $Q$ in Fig. \ref{fig:crossq}) 
%Here $S$ are the cross-sectional areas of the wires.
Also shown is the value of $I_s$ at the value $V$, obtained
by adding the integral of $ d I_s / dV$ to the equilibrium
value of $I_s$.  Note in particular that for large $V$,
the supercurrent has actually an opposite sign from
the equilibrium one, thus producing a "$\pi$-junction".
( {\it c.f.} \cite{volkov95,volkov97,buleavskii}) 

%%%%%%%%%%%%%%%%%%%%%%%%%%%%%%%%%%%%%%%%%%%%%%%%%%%%%%%%%%%%%%%%%

%of Fig. \ref{fig:cross} with 
%%%%%%%%%%%%%%%%%%%%%%%%%%%%%%%%%%%%%%%%%%%%%%%%%%%%%%%%%%%%%%%

\begin{figure}
\centerline{ 
	\epsfysize=0.4\textwidth 
\rotate[r]
{\epsfbox{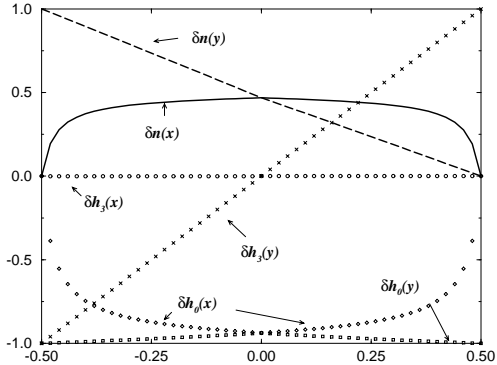} }
}
\vskip 0.5 cm
\begin{minipage}{0.45\textwidth}
\caption{ The distribution functions at $\epsilon = 0.24 E_D$
as functions of $x/L$ or $y/L$ 
for the cross geometry with parameters as in the last figure.
$\chi = \pi/4$.}
\label{fig:distr}
\end{minipage}
\end{figure}

To understand $d I_s / dV$, it is necessary to know
the behavior of the distribution functions for 
the quasiparticles.
(see Appendix for the technical details).
  I shall denote these functions on the
x- (y-) arms as $h_{0,3}(x)$ ($h_{0,3}(y)$) etc.
Since we are at $T=0$, a small change of the voltage at $V$
will affect only the occupation numbers at $ \epsilon = \pm e V$.
In Fig \ref{fig:distr} I have plotted the 
change of the distribution functions $\delta h_{0,3}$ at 
a relatively low energy when $V$ is increased from
below to above $e V = \epsilon$. At the S-reservoirs 
 $( x = \pm L_x/2) \ $ $\delta h_{0,3}= 0$ by 
choice, whereas at the normal reservoirs
 ( $y = \pm L_y/2$ ) $\delta h_3 \ = \pm 1$ and $\delta h_0 = -1 $.
The behavior of $\delta h_{0,3}$ is easy to understand
in this low energy limit,
where one can ignore the superflow ($Q$), the coupling between the
diffusion of the two distribution functions 
($M_{03}= - M_ {30}$ are small) and where the diffusivity for the 
particles ($\propto M_{33}$) reduces to that of the normal state.
Thus (see eq(\ref{h}) )
$\delta h_3 (y) $ is linear in $y$ and $\delta h_3 (x) \approx 0$.
Since there is an energy gap at the S-reservoir, the
effective diffusivity of the energy ($\propto M_{00}$) 
 is suppressed
near $ x = \pm L_x /2$.  Thus $\delta h_0$ only
has small gradients and hence $\delta h_0 \approx -1$
everywhere except near $ x = \pm L_x /2$.
In the language of the more familiar occupation number
$n (\epsilon) = ( 1 - (h_0(\epsilon) + h_3(\epsilon) ) ) /2$, 
in this $\epsilon \to 0$ limit $\delta n (y)$ is linear
in $y$ and thus $\delta n = 1/2$ at $ (x,y) = (0,0)$.
$\delta n(x)$ is almost constant and $\approx 1/2 $ 
near $x \approx 0$ and
only changes rapidly to $0$ near the S-reservoirs.
A finite $\epsilon$ provides a correction
to the above picture as can be also seen from Fig. \ref{fig:distr}.
\cite{notecor}
The values of $\delta n $ at $\epsilon = - eV$ can
be obtained by symmetry since
$n (- \epsilon) = ( 1  + h_0(\epsilon) - h_3(\epsilon) ) /2$. 
At this energy $\delta n(y)$ changes from $0$ at 
$y = - L_y/2$ to $-1$ at $y = L_y/2$
and $\delta n (x) \approx - 1/2$ near the center of the cross.

%%%%%%%%%%%%%%%%%%%%%%%%%%%%%%%%%%%%%%%%%%%%%%%%%%%%%%%%%%%%%%%%%

%%%%%%%%%%%%%%%%%%%%%%%%%%%%%%%%%%%%%%%%%%%%%%%%%%%%%%%%%%%%%%%

If $\delta n(x)$ were exactly $ \pm 1/2$ at $\epsilon = \pm eV$
and if one ignores the fact that $\delta n$ is actually
$x$ dependent, with eqn (\ref{current}) 
(or the equivalence of eqn (\ref{JsT}) )  for the 
current it is obvious that $d I_s / dV$ will be
equal to $-N_f D Q S$ at the corresponding energy.
In this case then
at large $V$ the current $I_s$ would be exactly zero.  
However, the actual current consists of both the 
supercurrent and the contributions from
the gradients of distribution functions. 
(see eq(\ref{current}) ).  Moreover $\delta n(x)$
is not exactly $\pm 1/2$ even at $x = 0$ when
$\epsilon$ is finite.
Thus the above approximation becomes worse
as the energy increases, making in general
the magnitude of $d I_s/dV$ somewhat smaller
than that of $N_f D Q S$. 
In particular the positive hump of $d I_s /dV$ 
at large $V$ (near $10 E_D$ in this particular example)
is smaller than the corresponding dip in $Q$
near that energy.  Hence  at large voltages $I_s$ becomes
negative as noted above.  \cite{noteequal}

%%%%%%%%%%%%%%%%%%%%%%%%%%%%%%%%%%%%%%%%%%%%%%%%%%%%%%%

In conclusion,  in this paper I studied the current-carrying 
density of states 
of a junction consisting of a dirty normal metal between
two superconductors.  I have also considered 
 the dependence of the 
supercurrent between the two superconductors
on the applied voltages at the normal reservoirs
of a cross geometry.

%%%%%%%%%%%%%%%%%%%%%%%%%%%%%%%%%%%%%%%%%%%%%%%%%%%%%%%%%%%%%

\vskip 0.5 cm

This research was supported by the NSF through the
the Science and Technology Center for Superconductivity, 
grant no.  DMR 91-20000, 
 Academy of Finland under  research grant No. 4385,
and the  {\AA}bo Akademi.

%%%%%%%%%%%%%%%%%%%%%%%%%%%%%%%%%%%%%%%%%%%%%%%%%%%%%%%%%%%%%%%%%%

\vskip 2 cm

{\bf Appendix}

In this appendix I summarize some basic equations for
easy reference. ({\it c.f.}, e.g. \cite{volkov94})
The basic equation to be solved is the Usadel equation

\begin{equation}
[\epsilon \tau_3, \check g] + 
 {D \over \pi} \partial_\mu (\check g  \partial_\mu \check g) = 0
\label{usadel}
\end{equation}
together with the normalization condition
$\check g ^2 = - \pi^2 \check 1 \ $ 
governing the angular averaged matrix Green's function 
$\check g$  
which in turn has  $\hat g^{R,A,K}$, the retarded, advanced, and Keldysh 
 martix Green's
functions as its components.
Here
$\epsilon$ is the energy.  The pair-breaking  mentioned in
the text is simulated  by $\epsilon \to \epsilon + i \gamma$
where $\gamma > 0$. \cite{notegamma}

$\hat g^R$ can be parameterized as 
$ - i \pi ( {\rm cos} \theta \tau_3 - 
   {\rm sin} \theta {\rm cos} \phi \sigma_2 \tau_1 
        + {\rm sin} \theta {\rm sin} \phi \sigma_2 \tau_2 )$.
$\hat g^A$ can be related to $\hat g^R$ by symmetry.
The variables $\theta$ and $\phi$ obey the differential
equations
\be
2 i ( \epsilon + i \gamma ) { \rm sin} \theta
  + 
D [ \partial_x^2 \theta -  {\rm sin} \theta {\rm cos} \theta
               (\partial \phi)^2  ] =0
\label{bulkt}
\end{equation}

and

\be
\partial ( {\rm sin}^2 \theta \partial \phi) = 0
\label{Iphi}
\ee

\noindent with the boundary conditions that they assume
their equilibrium values at the reservoirs.
For a normal reservoir $\theta = 0$, while 
at a superconducting reservoir 
$ {\rm cos} \theta = - i ( \epsilon + i \gamma) / {\cal D}$
where $ {\cal D} \equiv \sqrt{ \Delta^2  - ( \epsilon + i \gamma)^2} $.
($\gamma \to 0_+$ if pair-breaking is not included).

$Q$, related to the current-carrying density of states
as discussed in the text, is given by
\be
Q = 2 \ {\rm Im} [  {\rm sin}^2 \theta \partial \phi ]
\ee
It is thus then position independent within any wire
by eqn (\ref{Iphi}), a result which can also be directly obtained
from the definition (\ref{defQ}) for $Q$ and by taking the 
appropriate trace of eq (\ref{usadel}).
$Q$ obeys the symmetry $Q (-\epsilon) = -Q (\epsilon)$.

$\hat g^K$ is expressed via the distribution function
$\hat h$ as $\hat g^R \hat h - \hat h \hat g^A$ where
 $\hat h$ can be
chosen diagonal: $\hat h = h_0 \hat \tau_0 + h_3 \hat \tau_3$.
The distribution functions obey the equations 

\be
\partial [ Q h_0 +  ( M_{33} \partial h_3 + M_{30} \partial h_0) ] = 0
\label{h}
\ee
and the equation with $0 \leftrightarrow 3$. 
These two equations
express respectively the conservation of particle and energy
at each individual energy (due to the absence of interactions).
The (real) $M_{ij}$ coefficients are defined by
$ M_{ij} \equiv \delta_{ij} + {1 \over 4 \pi^2} {\rm Tr}
      [\hat g^A \tau_i \hat g^R \tau_j ]$.

The distribution functions at the reservoirs are given by
their equilibrium values.  At voltage $V$,
$h_0(\epsilon) = [ {\rm tanh} {\epsilon + e V \over 2 T} 
            + {\rm tanh} {\epsilon - e V \over 2 T} ] / 2$ and
$h_3(\epsilon) = [ {\rm tanh} {\epsilon - e V \over 2 T} - 
{\rm tanh} {\epsilon + e V \over 2 T} ] /2 $.  Thus,
at $T=0$, when the voltage sweeps through the corresponding
energy $\epsilon = e V$, the distribution functions at $y = - L_y/2 $
change by $\delta h_0 = - 1$ and $ \delta h_3 = -1$.
% (see Fig. \ref{fig:distr} ).
At the point where the voltage is  $ -V \ $ (  $y =  L_y/2 $ ), 
 $\delta h_0 = - 1$ and $ \delta h_3 = 1$.
 (see Fig. \ref{fig:distr} ).

The total number current density is given by
\be
J^N = {N_f D \over 2} \int d \epsilon 
  [ Q h_0 +  ( M_{33} \partial h_3 + M_{30} \partial h_0) ]
\label{current}
\ee
The three terms represent respectively the 
contributions from occupation of current-carrying states,
ordinary diffusion (with a  modified diffusion coefficient)
and an extra contribution due to broken particle-hole symmetry.

If a potential barrier exists, there will be discontinuites of the 
parameters $\theta$, $\phi$ across the barrier. 
The appropriate boundary conditions are derived from
\cite{kuprianov88}

\be
(2 N_f D S) \check g  \partial_\mu \check g
  = {1 \over 2 R_b}  [ \check g (x_{b-}), \check g (x_{b+}) ]
\ee
where $R_b$ is the resistance of the barrier at $x_b$.

\end{multicols}


\begin{thebibliography}{9}
\bibitem{magnitude} The magnitude of $Q$ will
be closer to the one without the side arms if we
increase the resistivity or length of the $y$ arms.
\bibitem{notecor}  This correction in particular 
becomes large if $\epsilon > \Delta$.
It is easy to see that, 
for the specific example given where
the x- and y- arms have identical resistivity
and length, then $\delta n( x \approx 0) \to \pm 1/4$
for $\epsilon >> \Delta$.
\bibitem{noteequal} 
From the arguments above it is clear that
for the approximate equality
between $d I_S /dV$ and
 $- N_f D Q S$ at low energies to hold, it is essential
that the two y-arms connecting the junction
to the N reservoirs are symmetric.
Neither $L_x = L_y$ nor  the equality
of the resistivity between the x- and y- arms
are necessary.  In particular, one can
for example decrease $I_y$ by increasing
the resistivity or the lengths of the y-arms,
though this will decrease the width of
of the energy region where the
above mentioned approximation remains good.
\bibitem{notegamma} This $\gamma$ which I have included is somewhat
different from magnetic scattering.  For magnetic
scattering the term that should be added is
actually $i \gamma {\rm cos} \theta$.
In this case a minigap $\epsilon_g$ will
still exist, where for energies below $\epsilon_g$, 
the Usadel equations allow solutions with
${\rm Re} \theta = \pi /2$, and ${\rm Im} \phi = 0$,
with $i \gamma {\rm cos} \theta$ real.
\end{thebibliography}
\end{document}